\documentclass[floatfix,twocolumn,aps,prl,showpacs,superscriptaddress]{revtex4-1}
\usepackage{graphicx}
\usepackage{amsmath}
\usepackage{amssymb}

\newcommand{\threevec}[3]{\left(\begin{array}{c}#1\\#2\\#3\end{array}\right)}
\newcommand{\inleva}[1]{\langle#1\rangle}
\newcommand{\abs}[1]{\left|#1\right|}
\newcommand{\nematic}{\mathbf{\hat{d}}}
\newcommand{\expF}{\langle\mathbf{\hat{F}}\rangle}
\newcommand{\absF}{\ensuremath{|\langle\mathbf{\hat{F}}\rangle|}}
\newcommand{\xhat}{\ensuremath{\mathbf{\hat{x}}}}
\newcommand{\yhat}{\ensuremath{\mathbf{\hat{y}}}}
\newcommand{\zhat}{\ensuremath{\mathbf{\hat{z}}}}
\newcommand{\rr}{\ensuremath{\mathbf{r}}}

\begin{document}

\author{Magnus O.\ Borgh}
\affiliation{Mathematical Sciences, University of Southampton, SO17 1BJ,
    Southampton, UK}

\author{Muneto Nitta}
\affiliation{
Department of Physics, and Research and Education Center for Natural
Sciences, Keio University, Hiyoshi 4-1-1, Yokohama, Kanagawa 223-8521, Japan}

\author{Janne Ruostekoski}
\affiliation{Mathematical Sciences, University of Southampton, SO17 1BJ,
    Southampton, UK}

\title{Stable core symmetries and confined textures for a vortex line in a spinor Bose-Einstein condensate}

\begin{abstract}
We show how a singly quantized vortex can exhibit energetically
stable defect cores with different symmetries in an atomic spin-1
polar Bose-Einstein condensate, and how a stable topologically
nontrivial Skyrmion texture of lower dimensionality can be confined
inside the core. The core isotropy and the stability of the confined
texture are sensitive to Zeeman level shifts. The observed structures
have analogies, respectively, in pressure-dependent symmetries of
superfluid liquid $^3$He vortices and in the models of superconducting
cosmic strings.
\end{abstract}

\date{\today}
\maketitle

Topological defects and textures cannot be removed from the system by
any continuous local transformation,
rendering them inherently robust and stable. Their topological
properties are universal throughout nature
from high-energy physics and cosmology to superfluids and liquid
crystals~\cite{pismen}. Defects and textures in different physical systems can
therefore frequently be characterized by generic effective theories
that are less sensitive to underlying microscopic interactions between
constituent particles.
Here the highly accessible atomic systems provide
unprecedented opportunities to act as
simulators~\cite{Blochreview,georgescu_rmp_2014} of phenomena in
condensed-matter systems, high-energy physics and
cosmology. For instance,
defect formation in nonequilibrium phase transitions using the
Kibble-Zurek mechanism~\cite{kibble_jpa_1976,zurek_nature_1985} has
been realized
experimentally in both scalar and spinor atomic Bose-Einstein Condensates
(BECs)~\cite{sadler_nature_2006,weiler_nature_2008,lamporesi_nphys_2013,navon_science_2015}
and the creation of black-hole analogs has been proposed
theoretically~\cite{garay_prl_2000}.
Studies of cosmological phenomena are also known in superfluid liquid
$^3$He~\cite{volovik,bauerle_nature_1996,ruutu_nature_1996}, despite
their less flexible experimental control and more indirect detection methods.

Recent experimental progress in the studies of spinor-BEC topological
defects and textures has culminated in the controlled
preparation~\cite{ray_nature_2014,ray_science_2015} of the atomic
superfluid analogs of
Dirac~\cite{savage_pra_2003,pietila_prl_2009_dirac,ruokokoski_pra_2011}
and 't~Hooft-Polyakov~\cite{stoof_prl_2001,ruostekoski_prl_2003}
monopoles, coreless
textures~\cite{leanhardt_prl_2003,leslie_prl_2009,choi_prl_2012,choi_njp_2012},
and  \emph{in situ} observation of the spontaneous breaking of core
axisymmetry
of a singly quantized vortex. In the last case the vortex with an
initially isotropic core split into
two half-quantum vortices~\cite{seo_prl_2015},
confirming the theoretical prediction~\cite{lovegrove_pra_2012}.

In this Letter, we demonstrate that the stable core
of a singly quantized spin-1 vortex, which was studied in the recent
experiments~\cite{seo_prl_2015}, can exhibit different axisymmetries
that continuously vary with imposed Zeeman shifts,
and that the vortex core
can host energetically stable, topologically
nontrivial confined spin textures.
The coexistence of energetically stable vortex cores
exhibiting different symmetries is analogous to the vortex cores of different
isotropies encountered in superfluid liquid
$^3$He~\cite{salomaa_rmp_1987,vollhardt-wolfle,kondo_prl_1991}.
The confined stable spin texture inside the vortex corresponds to a
continuous one-dimensional (1D) baby Skyrmion~\cite{supplemental}.

Defects or textures of lower dimensionality that are confined inside
the cores of singular host defects form elegant topological
objects. Here we show
that a 1D texture trapped inside the core of a singular vortex
line can be prepared and energetically stabilized
by engineering spatial profiles of the Zeeman shifts.
These fix the boundary conditions on the spin texture, ensuring that
it corresponds to a 1D baby Skyrmion with a nontrivial $1/2$
topological charge, forming an analogy to the Witten model of a scalar field
winding along a cosmic string defect line~\cite{witten_npb_1985}. A closed loop
of such a string is a cosmic vorton~\cite{radu_physrep_2008}. Although the
analogs of vortons and 3D Skyrmions~\cite{manton-sutcliffe} have been
actively investigated in atomic
BECs~\cite{ruostekoski_prl_2001,battye_prl_2002,savage_prl_2003,ruostekoski_pra_2004,kawakami_prl_2012,al-khawaja_nature_2001},
their experimental preparation has proved to be challenging due to
their inherently complex 3D structure---complications that could be
avoided by the simple preparation protocol of the confined 1D texture
we propose here.
Aside from their interest in cosmology, lower-dimensional Skyrmions
confined by domain walls or singular line 
defects appear also in high-energy physics~\cite{gudnason_prd_2014},
e.g., as the stable state of 
confined monopoles in quantum chromodynamics
(QCD)~\cite{tong_prd_2004,nitta_npb_2011,eto_jpa_2006,shifman_rmp_2007}.

In the mean-field theoretical model
the condensate wave function $\Psi$ is represented by the atomic
density $n=\Psi^\dagger\Psi$ and a three-component spinor $\zeta$, such that
\begin{equation}
  \label{eq:spinor-def}
  \Psi(\rr) = \sqrt{n(\rr)}\zeta(\rr) =
  \sqrt{n(\rr)}\threevec{\zeta_+(\rr)}{\zeta_0(\rr)}{\zeta_-(\rr)},
  \quad
  \zeta^\dagger\zeta = 1.
\end{equation}
The expectation value $\expF$ of the
spin operator gives the local condensate spin. The Hamiltonian
density is then~\cite{kawaguchi_physrep_2012}
\begin{equation}
  \label{eq:hamiltonian-density}
    {\cal H} =
    h_0
    + \frac{c_0}{2}n^2
    + \frac{c_2}{2}n^2|\mathbf{\inleva{\hat{F}}}|^2\\
    - pn\inleva{\hat{F}_z}
    + qn\inleva{\hat{F}_z^2}\,,
\end{equation}
where $h_0=(\hbar^2/2m)\abs{\nabla\Psi}^2 +
(m\omega^2/2)(x^2+y^2+z^2/4)n$, assuming a slightly prolate trap.
Linear and quadratic Zeeman shifts are represented by $p$ and $q$.
The interaction strengths are given by
$c_0=4\pi\hbar^2(2a_2+a_0)/3m$ and $c_2=4\pi\hbar^2(a_2-a_0)/3m$,
where $a_f$ is the $s$-wave scattering length in the spin-$f$ channel.

There are several theoretical studies of vortex structures in spin-1
BECs~\cite{yip_prl_1999,leonhardt_jetplett_2000,isoshima_pra_2002,kita_pra_2002,mizushima_prl_2002,martikainen_pra_2002,reijnders_pra_2004,mueller_pra_2004,saito_prl_2006,ji_prl_2008,takahashi_pra_2009,lovegrove_pra_2012,kobayashi_pra_2012,borgh_prl_2012,lovegrove_prl_2014}
that result from the rich order parameter space supported by the
Hamiltonian~\eqref{eq:hamiltonian-density}.
Here we assume $c_0,c_2>0$ (we take the value $c_0/c_2 \simeq 28$ for
$^{23}$Na~\cite{knoop_pra_2011}), in which case the interaction energy
favors the 
polar phase with $\absF=0$ (for a uniform system). The order parameter
$\zeta$ is then fully specified by the condensate phase $\tau$ and an
unoriented
unit vector $\nematic$~\cite{leonhardt_jetplett_2000,supplemental},
exhibiting \emph{nematic order}:
$\zeta(\tau,\nematic)=\zeta(\tau+\pi,-\nematic)$.
Vortices may then be characterized by the winding of both $\tau$ and
$\nematic$, where, however, only $\tau$ contributes to the
superfluid current and determines the topological charge.
The nematic order
leads to the existence of half-quantum
vortices where a $\pm\pi$ winding in $\tau$ is compensated by a
$\nematic\to-\nematic$ rotation~\cite{leonhardt_jetplett_2000}.
These were recently observed in experiment~\cite{seo_prl_2015}.  Note
that the name half-quantum vortex is sometimes used also in
two-component condensates, e.g., of
exciton-polaritons~\cite{rubo_prl_2007,lagoudakis_science_2009}, where,
however, the structure does not arise from nematic order as in
superfluid liquid $^3$He~\cite{salomaa_rmp_1987,vollhardt-wolfle},
liquid crystals~\cite{Kleman}, or atomic BECs, but is more 
reminiscent of a topologically very different coreless
vortex~\cite{matthews_prl_1999}. 
In the spin-1 BEC, the nematic order also allows several different ways of
forming a vortex with a given topological charge (see for example
Refs.~\cite{kawaguchi_physrep_2012,borgh_pra_2013} for detailed
presentations of the basic vortices).
Here we consider a singly quantized
vortex with an associated $2\pi$ winding of $\nematic$.
This may be written as
\begin{equation}
    \label{eq:012}
    \zeta = \frac{1}{\sqrt{2}}
       \threevec{-\sin\beta}
                {\sqrt{2}e^{i\varphi}\cos\beta}
                {e^{i2\varphi}\sin\beta},
\end{equation}
such that
$\nematic=\cos\varphi\sin\beta\xhat+\sin\varphi\sin\beta\yhat+\cos\beta\zhat$
and $\tau=\varphi$~\cite{leonhardt_jetplett_2000}, where $\varphi$ is
the azimuthal angle and $\beta$
the angle between $\nematic$ and the $z$ axis (taken to be constant).

We numerically minimize the mean-field
energy, Eq.~\eqref{eq:hamiltonian-density}, in the frame rotating at
frequency $\Omega$ (chosen to keep the vortex stable) around the $z$ axis, for
different core structures of the singly quantized polar vortex.
This is done by propagating the Gross-Pitaevskii
equations obtained from~\eqref{eq:hamiltonian-density} in imaginary
time using the split-step algorithm~\cite{javanainen_jpa_2006}.
We choose the nonlinearity $Nc_0=5000\hbar\omega\ell^3$ [$\ell
\equiv (\hbar/m\omega)^{1/2}$]. For
$^{23}$Na scattering lengths~\cite{knoop_pra_2011}, this yields
$N\omega^{1/2}\simeq7.5\times10^6$s$^{-1/2}$ between the number of
atoms and the trap frequency. For example, in a typical trap with
$\omega = 2\pi\times10$Hz, this corresponds to $N \simeq 10^6$ atoms.
We consider initial states constructed from Eq.~\eqref{eq:012},
and determine the
relaxed state in the presence of uniform and nonuniform Zeeman shifts
$p$ and $q$.

In addition to the Zeeman shifts arising from external magnetic fields,
an AC Stark shift corresponding to a highly tunable quadratic level shift
may be induced by combining a static magnetic field with a microwave
dressing field~\cite{gerbier_pra_2006}.
This method has
already been applied to the study of spin textures~\cite{guzman_pra_2011}.
Level shifts may also be induced by
lasers~\cite{santos_pra_2007}, which can allow for increased spatial control.

When $c_0>c_2$, it is energetically favorable for a polar vortex to
exhibit a core with nonzero superfluid density where
the BEC reaches the FM phase at the line
singularity~\cite{ruostekoski_prl_2003,lovegrove_pra_2012,kobayashi_pra_2012}.
For weak Zeeman shifts, this mechanism was predicted to lead to
breaking of axisymmetry in the vortex core by
splitting of the singly quantized vortex into a pair of
half-quantum vortices, exhibiting FM
cores whose spins anti-align, and preserving the
topology away from the core region~\cite{lovegrove_pra_2012,supplemental}.
The theoretical prediction was confirmed by the very recent
\emph{in situ} observation of the dissociation
process~\cite{seo_prl_2015}, where isotropic vortices were prepared by
means of trap and spin rotation.

\begin{figure}[tb]
  \vspace{0.2cm}
  \centering
  \includegraphics[width=\columnwidth]{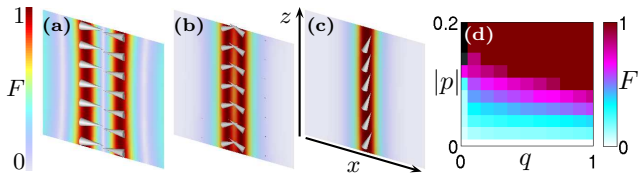}
  \caption{(Color online)
    (a)-(c) Spin vector
    (cones) and magnitude (color gradient) in stable vortex cores of
    different symmetries: 
    For weak Zeeman shifts, the broken axisymmetry of the vortex cores displays 
    half-quantum vortices with antialigning spins in the FM cores. The
    core symmetry is gradually restored as level shifts
    increase. Panels cover $|x|\leq3.9$, $|z|\leq3.6$ 
    (a) $p=q=0.0$, (b) $p=0.1\hbar\omega, q=0.4\hbar\omega$
    and (c) $p=0.1\hbar\omega, q=0.9\hbar\omega$.
    Trap rotation frequency $\Omega=0.26\omega$.
    (d) Spin magnitude
    $F=\absF$ at the local minimum between the singular lines
    ($F=1$ is the isotropic limit; black indicates pair not stable)
    as function of $p$ and $q$ (units of $\hbar\omega$).
  }
  \label{fig:axisymmetry}
\end{figure}

Here we find that the symmetry of the vortex core is sensitive to the
tunable Zeeman shifts. In particular, sufficiently 
large (spatially uniform) $p$ and $q$ may partially or completely
restore the axial symmetry of the vortex core.
This is illustrated in Fig.~\ref{fig:axisymmetry},
showing $\absF$ at its local minimum
between the singular lines, providing a measure for the degree of core
splitting, along with examples of stable vortex core structures exhibiting
different axisymmetry.
A positive quadratic shift favors population of the $\zeta_0$
component and acts to compress the core region.
However, this still
exhibits two distinct singular lines with antialigning spins, forcing
the wave function to the polar phase between them, also as $q$
increases.
If in addition $p$ is sufficiently large, the spins bend towards
the $z$ direction, and since the spins then do not antialign, a
contiguous nonpolar core region can form.
As $p$ and $q$ increase, the half-quantum vortex cores gradually
merge, forming a nonpolar core with broken axisymmetry, until the
splitting is suppressed entirely as $p\gtrsim0.1\hbar\omega$ and
$q\gtrsim0.5\hbar\omega$ [cf.\ Fig.~\ref{fig:axisymmetry}(d)],
leading to a single, axisymmetric FM core.
When $p$ approaches $0.2\hbar\omega$ and $q$ comes close to the trap energy,
the density in the vortex core is decreased, and the population of
$\zeta_-$ suppressed.

The existence of rich vortex core topologies has been predicted in
superfluid liquid $^3$He~\cite{salomaa_rmp_1987,vollhardt-wolfle}. Owing to the
flexibility of experimental 
preparation of atomic spinor gases and the atomic physics technology
for direct detection and control, the studies of vortex core
structures could be less challenging in atomic BECs. 
The coexistence of stable vortices with different
core symmetries in a spin-1 BEC is similar to that observed for
$B$-phase vortices in 
superfluid liquid $^3$He as the pressure is
varied~\cite{salomaa_prl_1983,salomaa_prl_1986,thuneberg_prl_1986,kondo_prl_1991}.
At high pressure, an axially symmetric vortex
core where the $A$ phase appears on the singular
line~\cite{salomaa_prl_1983} has the lower
energy~\cite{salomaa_prl_1986,thuneberg_prl_1986}.  However, at lower
pressure, energy is instead minimized by the ``double-core
vortex''~\cite{salomaa_prl_1986,thuneberg_prl_1986}, in which the
extended core region corresponds to two half-quantum vortices in the
planar phase.  In contrast to the continuously variable vortex-core symmetry in
the spinor BEC, the transition between the two core structures in
$^3$He is first
order~\cite{vollhardt-wolfle,salomaa_prl_1986,thuneberg_prl_1986}.

So far we have shown that spatially uniform Zeeman shifts can lead to
energetically 
stable vortex cores of different symmetries.
We now proceed to show that nontrivial, lower-dimensional textures
confined inside vortex 
cores can be engineered and
stabilized by designing Zeeman shifts with a
\emph{nonuniform} spatial profile.
In doing so, we consider again the singly quantized
vortex in Eq.~\eqref{eq:012}. However, we now make the additional
assumption that the
linear Zeeman shift has an engineered, nonuniform spatial dependence
$p(z)$, with different sign for positive and negative $z$. The Zeeman
energy strives to orient the spins in opposite directions in
the two parts of the vortex core.
For simplicity, we assume a linear gradient in $p$ between limits
$\pm|p_\mathrm{lim}|$ reached well within the extent of the gas. If
$|p_\mathrm{lim}|$ is sufficiently large, the spin bending energy may
be overcome and a nontrivial spin texture can be stabilized.
We further take the (spatially uniform) quadratic shift $q>0$
to be sufficiently large that the splitting
instability of the vortex is strongly suppressed.
The ground-state solution in the uniform system for the chosen parameters
is the polar
phase~\cite{zhang_njp_2003,ruostekoski_pra_2007,kawaguchi_physrep_2012}. 
We numerically minimize the energy in the rotating frame.

\begin{figure}[tb]
  \centering
  \includegraphics[width=\columnwidth]{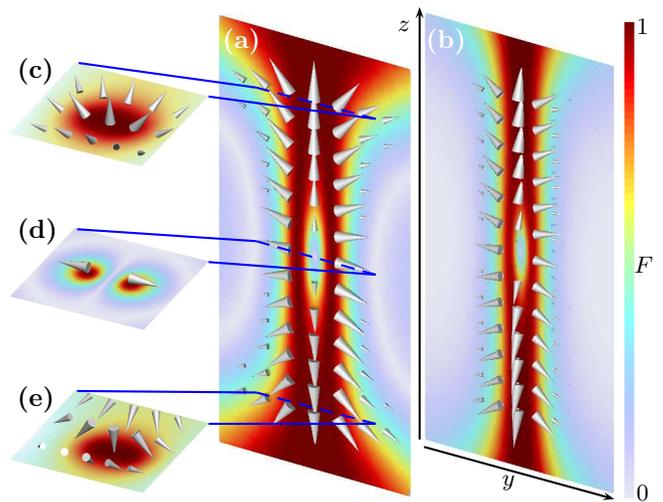}
  \caption{(Color online)
    (a)-(b) Spin vector
    (cones) and magnitude $F=\absF$ (color gradient) in the
    energetically stable, confined spin texture of a 1D baby Skyrmion.
    Boundary conditions are enforced by spatially nonuniform linear
    Zeeman shift,
    interpolating linearly between  $p=\pm0.2\hbar\omega$ over
    $|z| \leq 1\ell$ for $\Omega=0.25\omega$, with (a)
    $q=0.1\hbar\omega$ and (b) $q=0.4\hbar\omega$.
    The splitting instability of the vortex is suppressed,
    except in a small region around $z=0$. Panels cover $|y|\leq4.9$,
    $|z|\leq7.5$.
    (c)-(e) Cuts perpendicular to the vortex line, showing radial
    and cross disgyration connecting via the split region.
  }
  \label{fig:numerics012-210}
\end{figure}

A typical example of the resulting core structure is
shown in Fig.~\ref{fig:numerics012-210}, where an
energetically stable, continuous spin texture has emerged in the vortex
core.
Away from the
origin, the axial symmetry of the vortex core is not broken.
Around the origin, however, the vortex splits into two
separate cores over a short distance as $p$ switches sign.
The size of this region is insensitive to a very sharp $p$ gradient
but becomes more pronounced if the variation is slow compared with the
spin healing length $\xi_F=\hbar/(2m|c_2|n)^{1/2}$.
As the quadratic level shift $q$ is increased, the
unsplit vortex core of the singly quantized vortex becomes more
sharply defined and more axially symmetric, and the spin
texture consequently increasingly 1D.

\begin{figure}[tb]
  \centering
  \includegraphics[width=\columnwidth]{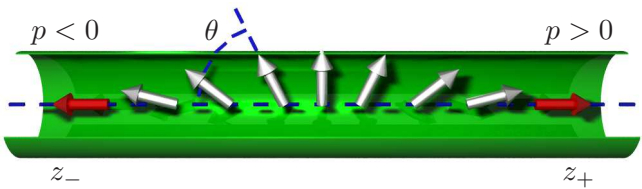}
  \caption{(Color online) Schematic illustration of the 1D
  half-Skyrmion spin texture (arrows) along a vortex core
  (cylinder). The angle $\theta$ between the spin vector and the
  vortex line plays the role of an $S^1$ order
  parameter. Red (dark gray) arrows indicate fixed
  boundary conditions leading to a $W=1/2$ 1D Skyrmion charge.
  }
  \label{fig:skyrmion-charge}
\end{figure}

The spin-vector field inside the vortex core relaxes
into a continuous texture where the condensate spin asymptotically
orients in
opposite directions for positive and negative $z$. (Note that the
spin texture in a plane perpendicular to the vortex line changes from
radial to cross disgyration across the origin. This is possible due to
the local splitting of the core.)
One may then think of the engineered linear Zeeman shift as imposing
fixed boundary conditions on the spin in the relaxed state due to an
energetic constraint. The FM
core provides an energetic confinement for the spin
texture, as $\absF$ quickly falls off towards zero outside it. The spin texture
along each singular line, where $\absF=1$, may then be viewed as a 1D
baby Skyrmion,
exhibiting a nonsingular, nontrivial winding of the spin vector
from $-\zhat$ to $\zhat$
(cf.\ schematic illustration in
Fig.~\ref{fig:skyrmion-charge}).

The 1D baby-Skyrmion texture constitutes
a lower-dimensional version of 3D Skyrmions, which
are 3D localized
particlelike textures~\cite{ruostekoski_prl_2001}, or 2D baby
Skyrmions, which are 2D fountainlike textures
(Anderson-Toulouse-Chechetkin coreless
vortices~\cite{chechetkin_jetp_1976,anderson_prl_1977}).
The winding of the 1D Skyrmion texture inside the vortex line is similar to
the winding of a scalar field along a
defect line in the Witten model of superconducting cosmic
strings~\cite{witten_npb_1985}.  Closed loops of such strings
constitute cosmic vortons, which are closely related to the 3D
Skyrmion~\cite{radu_physrep_2008}.
While attracting considerable theoretical interest, analogs of vortons
and Skyrmions have
proved difficult to realize in experiment due to the complicated 3D texture.
Here our proposed method for creating a 1D Skyrmion could provide an
experimentally simpler analog.

The 1D spin texture can be characterized by a winding number
\begin{equation}
  W = \frac{1}{2\pi}\int_{z_-}^{z_+} dz \frac{d\theta}{dz},
\end{equation}
where $\theta$ is the angle between the spin vector and the vortex line (see
Fig.~\ref{fig:skyrmion-charge}), and the integration interval covers
the whole texture.  $W$ is then the number of times the spin winds
around the corresponding order-parameter space $S^1$, and consequently takes an
integer value when the asymptotic spin vector is the same at both
ends of the texture. Whenever the boundary
conditions are fixed, for example by energetic constraints,
$W$ represents a conserved topological
charge.
Here the boundary conditions imposed by the Zeeman energy are
$\theta(z\to z_-)=0$ and $\theta(z\to z_+)=\pi$, and we may
thus ascribe a charge $W=1/2$ to the texture in
Fig.~\ref{fig:numerics012-210}.

Similar 1D baby-Skyrmion
textures along vortex lines appear in supersymmetric extensions of
QCD~\cite{gudnason_prd_2014}.
For example, a confined 1D Skyrmion texture forms the stable state of a confined
monopole~\cite{tong_prd_2004,nitta_npb_2011,eto_jpa_2006,shifman_rmp_2007}.
In this case a spin field $\mathbf{n}$ is confined to the
core of a vortex line. For the boundary conditions
$n_z(z\to+\infty)=+1$ and $n_z(z\to-\infty)=-1$, analogous to those
given by the Zeeman shifts in the atomic system, $\mathbf{n}$ takes
the form $n_z = \cos \theta, n_x= \sin \theta$, where
$\theta =  2 \arctan \{\exp [\sqrt{2} m (z-z_0)]\}$ is a sine-Gordon
kink solution with the kink at $z_0$.
The corresponding $\mathbf{n}$ texture is then similar to
Fig.~\ref{fig:skyrmion-charge} and to the confined-Skyrmion
spin texture found numerically for the spinor BEC in
Fig~\ref{fig:numerics012-210}.

Our results demonstrate that energy relaxation under
engineered Zeeman shift may
be used to experimentally realize the different core symmetries and
the confined baby-Skyrmion texture in the atomic spinor BEC.
The laser stirring in combination with a spin rotation used to generate
the singly quantized vortices in the recent core-splitting
experiment~\cite{seo_prl_2015} could be directly applied to create the
vortices.
Alternatively, singular vortices can be prepared
using phase-imprinting
techniques~\cite{matthews_prl_1999,leanhardt_prl_2002,shin_prl_2004,andersen_prl_2006,leslie_prl_2009}.
The core structure can subsequently be measured \emph{in situ}
by phase-contrast imaging, as used in the recent vortex
experiment~\cite{seo_prl_2015}, or detected by
ballistic expansion and separation of the spinor components.

In conclusion, in a system analogous to recent spinor
experiments~\cite{seo_prl_2015} we have demonstrated 
the existence of different vortex core topologies
and composite defects where a stable
lower-dimensional Skyrmion
texture is confined inside the core of a vortex line.
For a singular spin-1 vortex line the spontaneous breaking of the
axisymmetry of the vortex core was experimentally
observed~\cite{seo_prl_2015}. We found that the core isotropy is
sensitive to tunable Zeeman 
shifts leading to the coexistence of different stable vortex core
symmetries, where the broken symmetry could be continuously restored
by experimental control. This could open up possibilities for the
studies of rich vortex core topologies previously predicted for
superfluid liquid $^3$He~\cite{salomaa_rmp_1987,vollhardt-wolfle}.
Adding spatially nonuniform Zeeman shifts results in a stable
nonsingular 1D Skyrmion spin texture that is confined inside the
singular vortex line core.
Remarkably, this confined texture emerges from a continuous spin-1
condensate with only a single chemical potential everywhere in space.
The structure is analogous to the models of
superconducting cosmic strings, and is also reminiscent of confined
textures in QCD.
In the cosmological models a scalar field winding along a cosmic
string defect is considered to provide additional stability for the
string that could be further investigated in the atomic system. 
In the atomic superfluid a rapid rotation leads to a vortex lattice
with a Skyrmion spin texture trapped along each vortex line, which can
be used in studies of collective interactions between the
lower-dimensional confined textures. 
Moreover, when a spatially nonuniform Zeeman shift profile is combined with
phase transition dynamics involving spontaneous defect
formation~\cite{sadler_nature_2006,weiler_nature_2008,lamporesi_nphys_2013,navon_science_2015},
we can also envisage simplified simulation models for superconducting
cosmic string formation scenarios, where the Kibble-Zurek mechanism
would involve an internal defect structure. 

\begin{acknowledgments}
We acknowledge financial support from the EPSRC.
The work of M.~N.~is supported in part by a Grant-in-Aid for
Scientific Research on Innovative Areas ``Topological Materials
Science'' (KAKENHI Grant No.~15H05855) and ``Nuclear Matter in Neutron
Stars Investigated by Experiments and Astronomical Observations''
(KAKENHI Grant No.~15H00841) from the the Ministry of Education,
Culture, Sports, Science (MEXT) of Japan, by the Japan Society for
the Promotion of Science 
(JSPS) Grant-in-Aid for Scientific Research (KAKENHI Grant
No.~25400268), and by the MEXT-Supported Program for the Strategic
Research Foundation at Private Universities ``Topological Science''
(Grant No.~S1511006).
We acknowledge the use of the IRIDIS High Performance
Computing Facility at the University
of Southampton.
M.~N.\ thanks the University of Southampton for
warm hospitality during which this project took shape.
\end{acknowledgments}

\appendix

\setcounter{equation}{0}
\setcounter{figure}{0}
\renewcommand{\theequation}{S\arabic{equation}}
\renewcommand{\thefigure}{S\arabic{figure}}

\section{Supplemental Material}

\noindent In this Supplemental Material we provide
additional discussion of the breaking of axisymmetry in the core of
a stable singly quantized vortex in the polar phase of the
spin-1 BEC. We also give a brief
overview of the Skyrmion textures in different dimensions.

\section*{Breaking of vortex-core axisymmetry}

\noindent
In the main text, we show that the singly
quantized vortex without internal structure can exhibit different,
energetically stable core
symmetries as the (spatially uniform) Zeeman shifts are varied.
When the level shifts are
weak, energy
relaxation leads to spontaneous breaking of axial symmetry in the
core, splitting the vortex into two half-quantum
vortices, as predicted in Ref.~\cite{lovegrove_pra_2012}. This
splitting of the vortex core was recently experimentally observed 
for a $^{23}$Na spin-1 BEC of $3.5\times10^6$
atoms in an oblate trap with $(\omega_x,\omega_y,\omega_z) = 2\pi
\times (4.2,5.3,480)$~Hz~\cite{seo_prl_2015}. In this experiment,
singly quantized 
vortices were created in a condensate initially occupying only the
$m=0$ Zeeman level.  A spin rotation is then applied by tuning the
quadratic Zeeman shift (induced using microwave
dressing~\cite{gerbier_pra_2006}) to transfer 
population to the $m=\pm1$ levels.  After the spin rotation,
the vortices are observed to split into half-quantum vortex pairs with
opposite core spin polarization. The resulting half-quantum vortices
were identified by \emph{in situ} imaging~\cite{seo_prl_2015}, in
which spin-dependent phase-contrast imaging is used to map out the
condensate magnetization.  The oppositely magnetized FM cores of the
half-quantum vortices can then be discerned.

The breaking of axisymmetry and splitting of the singly quantized
vortex is made possible by the
(uniaxial) \emph{nematic order} exhibited by the polar phase of the
spin-1 BEC, which allows the existence of half-quantum vortices.
The polar order parameter
may be expressed in terms of a condensate phase $\tau$ and a unit
vector $\nematic$ as~\cite{leonhardt_jetplett_2000}
\begin{equation}
  \zeta = \frac{e^{i\tau}}{\sqrt{2}}\threevec{d_x+id_y}{\sqrt{2}d_z}{d_x-id_y}
  = \frac{e^{i\tau}}{\sqrt{2}}\threevec{-e^{-i\alpha}\sin\beta}
                                       {\sqrt{2}\cos\beta}
				       {e^{i\alpha}\sin\beta}.
\end{equation}
(In the last expression $\nematic$ has been parametrized in terms
of azimuthal and polar angles $\alpha$ and $\beta$, for later
convenience). Note that
$\zeta(\tau,\nematic)=\zeta(\tau+\pi,-\nematic)$. These two states
must therefore be identified, and the vector $\nematic$ is understood
as an unoriented \emph{nematic axis}.  Rotations of $\nematic$ do not
contribute to the superfluid flow, making it possible to form a vortex
carrying half a quantum of circulation by letting a $\pi$ winding of
$\tau$ around the vortex line be accompanied by a $\nematic \to
-\nematic$ rotation of the nematic axis, keeping the order parameter
single-valued. For example, taking $d_z=0$ for simplicity, a
half-quantum vortex can be written as
\begin{equation}
  \label{eq:half-quantum}
  \zeta =
  \frac{e^{i\varphi/2}}{\sqrt{2}}\threevec{-e^{-i\varphi/2}}
                                          {0}
				          {e^{i\varphi/2}}
  = \frac{1}{\sqrt{2}}\threevec{-1}{0}{e^{i\varphi}},
\end{equation}
where $\varphi$ is the azimuthal coordinate around the vortex
line. Nematic order also gives rise to half-quantum vortices in,
e.g., the $A$ phase of superfluid liquid
$^3$He~\cite{vollhardt-wolfle}, and to $\pi$-disclinations in nematic
liquid crystals~\cite{Kleman}.
The name is also sometimes used in the
context of exciton-polariton condensates in reference to a vortex with
a $\pi$ rotation of linear polarization of the photon
component~\cite{rubo_prl_2007,lagoudakis_science_2009}. This does not,
however, arise from
nematic order, but is more reminiscent of a topologically very different coreless vortex in a
two-component BEC~\cite{matthews_prl_1999}.

We can now understand the splitting of the singly quantized vortex
given by Eq.~(3) of the main text as
\begin{equation}
  \label{eq:splitting}
  \begin{split}
    &\frac{e^{i\varphi}}{\sqrt{2}}
       \threevec{-e^{-i\varphi}\sin\beta}
                {\sqrt{2}\cos\beta}
                {e^{i\varphi}\sin\beta}
   \to\\
   &\frac{e^{i\varphi_1/2}}{\sqrt{2}}
       \threevec{-e^{-i\varphi_1/2}\sin\beta}
                {\sqrt{2}\cos\beta}
                {e^{i\varphi_1/2}\sin\beta}
  \oplus
  \frac{e^{i\varphi_2/2}}{\sqrt{2}}
       \threevec{-e^{-i\varphi_2/2}\sin\beta}
                {\sqrt{2}\cos\beta}
                {e^{i\varphi_2/2}\sin\beta}.
  \end{split}
\end{equation}
Here the spinors on the right-hand side represent half-quantum
vortices [cf.\ Eq.~\eqref{eq:half-quantum}].
In these, $\varphi_{1,2}$ are the azimuthal angles relative to each vortex
line, and the spinors describe the wave function
locally around each vortex core. Away from the core
region, the wave function still corresponds to the original singly
quantized vortex.
(Note that $\oplus$ here indicates the addition of topological defects.)

\section*{Skyrmions and baby Skyrmions}
\noindent Nonsingular textures may be
topologically nontrivial by considering maps from a compactified
real space to the compact order-parameter space. When the
order parameter reaches the same value everywhere sufficiently far away from the
(particlelike) texture, the entire boundary enclosing the texture may be
identified and the volume in $\mathbb{R}^3$
becomes topologically $S^3$ (a unit sphere in four dimensions).
One may then think of the $S^3 \to S^3$ map as distributing (an
integer number of copies of) the full order-parameter space over the
compactified real space. The corresponding nontrivial textures are the
3D Skyrmions~\cite{skyrme_1961}. Analogous structures may be constructed
in a two-component BEC~\cite{ruostekoski_prl_2001}.

An $S^2$ order-parameter space, 
may similarly be
distributed over a (2D) real-space surface with fixed, uniform
boundary conditions (corresponding to an $S^2 \to S^2$ map). Such a 2D Skyrmion
is commonly referred to as a (2D) ``baby Skyrmion'', being the
topologically lower-dimensional analog of the full 3D Skyrmion, and
may be realized as a coreless
vortex~\cite{leanhardt_prl_2003,leslie_prl_2009,choi_prl_2012,choi_njp_2012}. 
The dimensionality of the baby Skyrmion may be further reduced by
considering an $S^1$ order parameter. 
For uniform boundary conditions such that 1D space can be compactified to $S^1$,
the resulting $S^1 \to S^1$ map defines a 1D baby Skyrmion.
In our system a ferromagnetic spin texture confined inside the core of
a vortex line 
exhibits fixed boundary conditions. As the boundary conditions are
twisted (the orientation of the spin 
vector differs by $\pi$ in the two ends of the vortex line, the 1D
Skyrmion winding number is equal to 1/2. 
Any further winding of the spin texture would lead to higher Skyrmion
winding numbers.

\end{document}